
\def\ref{\par\noindent\hangindent=2pc \hangafter=1 }

\def\etal{{\it et al.~}}

\def\ee #1 {\times 10^{#1}}
\def\ut #1 #2 {\hbox{\thinspace #1}^{#2}}
\def\u #1 {\hbox{\thinspace #1}}
\def\msol{\hbox{$\hbox{M}_\odot$}}

\def\mdotos #1 {\left({\dot M \over 3\ee -3 \u {\msol} \ut yr -1
}\right)^{#1}}
\def\vwos #1 {\left({v_w \over 700 \u km \ut s -1 } \right )^{#1}}
\def\ros #1 {\left( {r \over 1 \u pc } \right )^{#1}}
\def\nos #1 {\left ( {n_{\rm H} \over 10^4 \ut cm -3 } \right)^{#1}}
\def\naos #1 {\left ( {n_a \over 10^4 \ut cm -3 } \right)^{#1}}
\def\nios #1 {\left ( {n_i \over 30 \ut cm -3 } \right)^{#1}}
\def\nhos #1 {\left ( {n_{\rm H} \over 10^5 \ut cm -3 } \right)^{#1}}
\def\tos #1 {\left ( {T \over 10^4 \u K } \right )^{#1}}
\def\taos #1 {\left ( {T_a \over 10^4 \u K } \right )^{#1}}
\def\bos #1 {\left({B\over 1\u mG }\right)^{#1}}
\def\vos #1 {\left ( {v \over 200 \u km \ut s -1 } \right )^{#1}}
\def\vaos #1 {\left ( {v \over 200 \u km \ut s -1 } \right )^{#1}}
\def\figpage #1 #2 {\vfill\eject\pageno=#1\vglue 8.0in\line{\hfill #2}}

\def\mathnew{\mathsurround=0pt}
\def\simov#1#2{\lower .5pt\vbox{\baselineskip0pt \lineskip-.5pt
\ialign{$\mathnew#1\hfil##\hfil$\crcr#2\crcr\sim\crcr}}}

\font\twelverm=cmr10 scaled 1200
\font\ninerm=cmr7 scaled 1200
\font\sevenrm=cmr5 scaled 1200
\font\twelvei=cmmi10 scaled 1200
\font\ninei=cmmi7 scaled 1200
\font\seveni=cmmi5 scaled 1200
\font\twelvesy=cmsy10 scaled 1200
\font\ninesy=cmsy7 scaled 1200
\font\sevensy=cmsy5 scaled 1200
\font\twelveex=cmex10 scaled 1200
\font\twelvebf=cmbx10 scaled 1200
\font\ninebf=cmbx7 scaled 1200
\font\sevenbf=cmbx5 scaled 1200
\font\twelveit=cmti10 scaled 1200
\font\twelvesl=cmsl10 scaled 1200
\font\twelvett=cmtt10 scaled 1200
\skewchar\twelvei='177 \skewchar\ninei='177 \skewchar\seveni='177
\skewchar\twelvesy='60 \skewchar\ninesy='60 \skewchar\sevensy='60
\def\twelvepoint{\def\rm{\fam0 \twelverm}
  \textfont0=\twelverm \scriptfont0=\ninerm \scriptscriptfont0=\sevenrm
  \rm
  \textfont1=\twelvei \scriptfont1=\ninei \scriptscriptfont1=\seveni
  \def\mit{\fam1 } \def\oldstyle{\fam1 \twelvei}
  \textfont2=\twelvesy \scriptfont2=\ninesy \scriptscriptfont2=\sevensy
  \def\cal{\fam2 }
  \textfont3=\twelveex \scriptfont3=\twelveex \scriptscriptfont3=\twelveex
  \textfont\itfam=\twelveit \def\it{\fam\itfam\twelveit}
  \textfont\slfam=\twelvesl \def\sl{\fam\slfam\twelvesl}
  \textfont\bffam=\twelvebf \scriptfont\bffam=\ninebf
    \scriptscriptfont\bffam=\sevenbf \def\bf{\fam\bffam\twelvebf}
  \textfont\ttfam=\twelvett \def\tt{\fam\ttfam\twelvett}
  }
\def\folio{\ifnum\pageno=1\nopagenumbers\else\number\pageno\fi}
\font\simlessrgebf=cmbx10 scaled 1440
\font\twelvess=cmss10 scaled 1200

\def\doublespace{\baselineskip 22.76 pt}
\hsize 6.5 true in
\vsize 8.7in
\hoffset=0. true in
\voffset=0. true in
\twelvepoint
\doublespace
\null
\centerline{Submitted to the Editor of the Astrophysical Journal {\it
Letters}  on December 13, 1993}
\null\vskip 0.85 true in
\centerline{\simlessrgebf Anisotropy in the Angular Broadening }
\centerline{\simlessrgebf  of Sgr A$^*$ at the Galactic Center}

\vskip 1.0 true in
\centerline{\bf Farhad Yusef-Zadeh}
\vskip 0.05in
\centerline{\sl Department of Physics and Astronomy, Northwestern University}
\vskip 0.05in
\centerline{{\bf William Cotton}}
\vskip 0.05in
\centerline{\sl National Radio Astronomy Observatory, Charlottesville}
\vskip 0.05in
\centerline{{\bf Mark Wardle}}
\vskip 0.05in
\centerline{\sl Department of Physics and Astronomy, University of Rochester}
\vskip 0.05in
\centerline{{\bf Fulvio Melia}\footnote{\hbox{$\null^1$}}{Presidential Young
Investigator.}}
\vskip 0.05in
\centerline{\sl Department of Physics and Steward Observatory,
University of Arizona}
\vskip 0.05in
\centerline{and}
\vskip 0.05in
\centerline{\bf Douglas A. Roberts}
\centerline{\sl Department of Astronomy, University of Illinois}

\vskip 0.05in
\vskip 0.2in
\vfill\eject
\centerline{\bf ABSTRACT}
\bigskip
{\twelvess
We present the results of a $\lambda$20 cm VLA\footnote{$^2$}{VLA is a
component of the National Radio Astronomy Observatory, which is
operated by Associated Universities, Inc., under contract to the
National Science Foundation} observation of the compact Galactic center
radio source Sgr A$^*$.  The scatter-broadened image is elongated in
the East-West direction, with an axial ratio of 0.6$\pm$0.05 and a
position angle of 87$^0\pm$3$^0$.  A similar shape and orientation has
been found previously at shorter wavelengths using VLBI and  VLBA.
Both the major and minor axes follow the $\lambda^2$ law  appropriate
for scattering by turbulence in the intervening medium.

Assuming that the anisotropy is caused by a magnetic field
permeating the scattering medium, we argue that the scattering occurs
within extended HII regions lying in the central 100 pc of the Galaxy.
The magnetic field in this region must be poloidal, organized and is
estimated to have a strength of at least 30 to 100 $\mu$Gauss.
}

\bigskip\noindent {\it Subject headings}:  galaxies:  ISM---
Galaxy: center ---ISM: individual (Sgr A$^*$, IRS 16)

\centerline{\bf 1. Introduction}
\medskip

Sgr A$^*$, the compact radio source at the Galactic center, shows nonthermal
characteristics with a spectrum that rises as $\nu^{0.25}$ at radio
wavelengths, and whose brightness temperature exceeds $10^{10}$K at
$\lambda$7mm (Backer 1993; Backer \etal 1993). With a radio luminosity of
$\sim 2 \times 10^{34}$ erg/s ($\sim 1.1$ Jy at $\lambda$2cm), Sgr A$^*$ is
the brightest radio source within the inner several degrees of the Galactic
center region. It lies close to the dynamical center of the Galaxy
and the surface density distribution of the stellar cluster that
engulfs the
Galactic center is centered to within roughly 0.1$"$ of  Sgr A$^*$
(Eckart \etal 1993).
Theoretical analysis of the radiation
 spectrum of this source is also consistent
with it  being a massive ($\approx$1-2 $\msol$) black hole (Melia 1992).
These properties as well as the intrinsic time variability (Brown and Lo 1982;
Zhao \etal 1991; Backer 1993) suggest that Sgr A$^*$ resembles the cores of
extragalactic radio sources, perhaps hinting at a common heritage.

The angular broadening of Sgr A$^*$ was first reported to scale
 quadratically with wavelength
$\lambda$ (Davies \etal 1976)  not long after  the discovery of
Sgr A$^*$ by Balick and Brown (1974).
 The broadening is caused by electron density
fluctuations which change the refractive index of the interstellar medium
along the line of sight. Anisotropies in the scatter-broadened image have
been reported at relatively high frequencies for some time (Davies \etal
1976;
Lo \etal 1985; Jauncey \etal 1989; Marcaide \etal 1992; Alberdi \etal 1993;
Lo \etal 1993; Krichbaum \etal 1993; Backer \etal 1993). In one of the
more recent measurements by Lo \etal (1993), the
power law index of the
major  axis is estimated to scale as $\lambda^{1.94\pm0.03}$.
Here we report VLA
observations showing similar results to those of
earlier VLBA and VLBI high-frequency measurements.
The new observation indicates  that the size of Sgr A$^*$ at $\lambda$20cm is
asymmetric in the East-West direction. The  measured size of
Sgr A$^*$ scales as $\lambda^{2.01\pm0.02}$ and $\lambda^{2.13\pm0.04}$
for the semi-major and semi-minor axes, respectively. We argue that the
turbulent medium responsible for scattering lies within the inner 100 pc of the
Galaxy and that the  asymmetry in the scatter-brodened size of
Sgr A$^*$ is accounted for by
an organized, large-scale
 poloidal magnetic field permeating the central region of
the Galaxy.
Estimates of the orientation and strength of the magnetic fields
made here
are quite consistent  with an  independent set of
earlier magnetic field
measurements  carried out toward
the nonthermal filamentary structures near the Galactic
center (Yusef-Zadeh, Morris \& Chance 1984; Sofue \etal 1987;
Yusef-Zadeh \& Morris 1987a).

\medskip
\centerline{\bf 2. Data Reductions and Results}

Radio continuum observations of the Galactic center were carried out on
January 10, 1985 using the A-array configuration of the VLA. We used
the spectral line mode of the VLA correlator by choosing 16 channels
of 1.5625 MHz
bandwidth with only one sense of polarization. This observation was carried
out with an antenna pointing  centered at
$\alpha(1950)=17^h 42^m 25.0^s, \delta(1950)=-28^0 57' 30''$. This
is offset by about 108.6$"$ to the north and 4.3$^s$ to the west of the
position of Sgr A$^*$.
The bandwidth smearing of Sgr A$^*$ due to individual channels is
roughly 140 mas at a PA$\approx -28^0$ which is smaller than the
scatter-broadened  size of Sgr A$^*$.  A correction for this effect was
made as described below.
The  $\lambda$20cm observations consisted of two 25-minute scans
separated in time by one hour.

 The channel data  were calibrated
by using 1748-253 and 3C286 as the phase and flux calibrators,
 respectively. In order to improve the dynamic
range of the final images, we also used
3C286 and 1748-253 as  bandpass calibrators before
the first 14 channels of the data  were averaged to form a continuum
data set.  The standard calibration procedures gave a flux density of
1.2075$\pm$0.02 Jy for 1748-253.  We employed
phase  self-calibration of  Sgr A$^*$ by
restricting the {\it uv} range to $>$20 k$\lambda$.
A first order correction for the effects of bandwidth smearing were
made using the AIPS program UVADC which corrects the amplitude loss in
each visibility based on a model of the emission in the field.
The final image was constructed after bandwidth smearing correction
was applied using only baselines longer than 20 k$\lambda$.
Due to the southerly declination of the source, the FWHM of the
synthesized beam is $2400\times1000$ mas with PA=-17$^0$.
There was 696 mJy of flux CLEANed from the image leaving the rms noise in a
blank region of about 1 mJy.  The region CLEANed was
$250''\times250''$ centered on the position of Sgr A$^*$.
The CLEANing was done using task WFCLN.

An elliptical Gaussian fit to the final CLEAN image of Sgr A$^*$ yields
a FWHM=$ 636\pm8\times 382\pm32$ mas and a position angle 87$^0\pm3^0$.
 In the {\it uv} domain, we also attempted
an elliptical Gaussian fit at long, $>$50k$\lambda$, {\it uv} spacings.
In spite of the extended structure in the field, the Gaussian  fits to
the {\it uv} data  gave FWHM=$680\pm7\times 459\pm3$ mas and a
position angle 96$^0\pm2^0$.
The axial ratio in the {\it uv} and in the image domain are
0.68$\pm0.005$ and 0.60$\pm0.05$, respectively.
The {\it uv} plane results do not appear to be terribly sensitive to
the initial guess; however, using shorter baselines tends to make the
source rounder.
We believe that the result of Gaussian fitting in the image
plane is more reliable than in the {\it uv} plane since the
$\lambda$20cm image of the field shows  100 mJy of flux outside of Sgr
A$^*$ even excluding the shorter baselines. The parameters of the fit
stated here are also more accurate than those given in recent
preliminary presentation of the $\lambda$20cm data (Yusef-Zadeh 1993).

Jauncey \etal (1989)  found an asymmetry at $\lambda$3.6cm and was
recently  confirmed by Lo \etal (1993). A similar asymmetry
with an axial ratio close to 0.5$\pm$0.2 is also noted at 1.2cm (Marcaide
\etal 1992; Lo \etal 1993),
 although recent MK II VLBA measurements at $\lambda$1.2cm gave
a size of about 2.8 mas and were  not able to constrain the axial ratio at
this critical wavelength (Walker, private communication). Backer \etal
(1993) have recently reported the best elliptical Gaussian fit model to the
closure amplitude data at $\lambda$7mm, which gives a size
0.74$\pm0.03\times0.04\pm0.20$ mas with the position angle 90$^0\pm10^0$. The
elongation of the source at this wavelength is consistent with the
$\lambda$7mm data presented by Krichbaum \etal (1993).
Table 1 shows a list of position angle and axial ratio measurements of
Sgr A$^*$ at a number of wavelengths. (Krichbaum \etal find
Gaussian fits to two sources and  the parameters of the fit to the
strong source is shown in Table 1).
Considering the uncertainties involved in measuring the minor axis of
Sgr A$^*$ at high frequencies, the axial ratio of 0.6 at $\lambda$20cm
presented in this paper appears to be consistent with  the previous VLB
measurements at $\lambda$3.6 and 1.35cm (Jauncey \etal 1989; Lo \etal
1993; Alberdi \etal 1993).

\medskip
$$\vbox{\offinterlineskip
\halign{
\strut \quad $#$ \hfil
&\quad \vrule \hfil $#$ \hfil
&\vrule\quad \hfil $#$ \hfil
&\quad \vrule\quad \hfil $#$ \hfil
&\quad \vrule\quad # \hfil \cr
\lambda \u (cm) & \hbox{\ Major axis (mas)\ } & \hbox{Axial ratio}
& \hbox{PA ($\,^{\rm o}$) } & \hbox{Reference} \cr
\noalign{\hrule}
\noalign{\hrule}
0.7   &  0.74\pm0.03  &  0.54\pm0.2   &   90\pm10  & Backer \etal 1993 \cr
0.7   &  0.70\pm0.1   &   0.6    &  -65\pm20  & Krichbaum \etal 1993 \cr
1.35  &   2.6\pm0.2   &   0.5\pm0.2   &   75\pm10  & Lo \etal 1993 \cr
1.35  &  2.58\pm0.08  &   0.5\pm0.2   &   79\pm6   & Alberdi \etal 1992 \cr
3.6   &  17.4\pm0.5   &  0.53\pm0.1  &   82\pm6   & Jauncey \etal 1989 \cr
3.6   &  17.5\pm0.5   &  0.49\pm0.06  &   87\pm5   & Lo \etal 1993 \cr
20.7  &   636\pm8     &  0.60\pm0.05  &   87\pm3   & this paper \cr
}
}$$
\medskip
\centerline{\bf Table 1}
\medskip

The position angle of the asymmetry in the scattered-broadened image of Sgr
A$^*$, as seen in Table 1,
 is consistent with the East-West elongation as noted at $\lambda$3.6,
1.2cm and 0.7cm. An axial ratio of 0.6 at $\lambda$20cm
confirms that the
intervening scattering medium is anisotropic. There is
a hint that the shape of Sgr A$^*$ becomes more anisotropic at shorter
wavelengths, though this needs be confirmed in future observations.
Figure 1 shows the relationship between the size of Sgr A$^*$
vs. $\lambda$ for the semi-major and semi-minor axes
as a function of wavelength. Also shown is a least-square fit to the data,
 where
the mean of the logarithmic upper and lower error bars has been used to
determine the weighting for each data point. The major axis follows a slope
-2.01$\pm0.02$ whereas the fit to the size of the minor axis is steeper,
-2.13$\pm0.04$ as it passes through the recent $\lambda$7mm data point
(Backer \etal 1993). The slope of the size of the minor axis is dictated by
two data points at $\lambda$20 and 3.6cm, though consistent with
high-frequency points with large error bars. Obviously, future VLBA
observations at a number of frequencies,  should be able to determine if the
size of Sgr A$^*$ become somewhat more asymmetrical at very high frequencies.

\medskip
\centerline{\bf 3. Discussion}

It has been suggested (Jauncey \etal 1989; Lo \etal 1993; Backer \etal
1993) that the elongation of Sgr A$^*$
is caused by anisotropy of the turbulence in the scattering medium
imposed by a strong magnetic field, which permits small-scale density
and velocity fluctuations perpendicular to the direction of the field
(Cordes \etal 1984;
 Higdon 1984, 1986).  The scattered image is stretched in a direction
perpendicular to the average projection of the field in the plane of
the sky, so the inferred average field direction is oriented
North-South.

Van Langevelde \etal (1992) found that OH/IR stars within 15 arcminutes
of the Galactic Center suffer a large angular broadening.  They
suggested that the scattering medium could either lie close to the
Galactic center or could be relatively local.  The observed anisotropy
favours the former model because the inferred magnetic field direction
is consistent with the large scale poloidal field that exists in the
central 100 to 200 parsecs of the Galaxy (Yusef-Zadeh, Morris and
Chance 1984; Tsuboi \etal 1986) rather than with the field direction
of the Galactic disk.  Assuming that the position angle of
the scatter-broadened shape of  Sgr A$^*$ is 80$^0\pm10^0$,
the predominant component of the
magnetic field is inferred to be oriented
 70$^0\pm10^0$ away from the Galactic plane.

A survey by Mezger and Pauls (1979) shows that the inner 150 pc
$\times$ 50 pc (l$\times$b) of the Galaxy is enveloped in strong radio
continuum emission.  Half this emission is now known to be thermal,
arising from extended HII regions (Schmidt 1978; Handa \etal 1987).
These HII
regions are good candidates for the source of scattering since they can
support anisotropic turbulence on small lengthscales (Higdon 1984).  As
a simple model for the scattering medium we assume that HII regions are
uniformly distributed within the inner 100 pc of the Galaxy.  The HII
regions have characteristic electron density $n_e$, with rms
fluctuation $\delta n_e$, and the total path length that intercepts the
HII regions is $L$.
{}From Mezger and Pauls (1979), the emission measure due to thermal
ionized gas is
$n_e^2 L \approx 2\ee 4 \ut cm -6 \u pc $, whereas the observed
angular size of Sgr A* requires that $\delta n_e^2 L \approx 200 \ut cm
-6 \u pc $ (Backer 1978).  This implies that $\delta n_e/n_e \approx
0.1$, independent of the degree of clumping of the ionized gas.

A lower limit can be placed on the magnetic field in this region because
it has to be strong enough to resist being bent by turbulence, that
is, the field pressure should be at least as large as the thermal
pressure. Adopting a temperature of $10^4 \u K $, this yields
$B/(n_e)^{1/2} > 8 \ut cm 3/2 \,\mu {\rm G} $.  Thus for $L=100\u pc $,
there is one extended HII region with $n_e = 17 \ut cm -3 $ and $B > 30 \,
\mu \u G $ whereas for $L=1 \u pc $, we obtain $n_e = 170 \ut cm -3 $, and $B >
100 \, \mu$G.  Such a high value of the magnetic field is not
unexpected since a number of large-scale magnetized filaments in the
Galactic center region are thought to have milliGauss field strengths in the
direction perpendicular to the Galactic plane (Yusef-Zadeh
\& Morris 1987b; Serabyn \& G\"usten 1991).

The field strength can be independently estimated from Faraday rotation
measurements.  The observed rotation measure ranges between 1000 and
5500 rad. $\ut m -2 $ (Inoue \etal 1984;
Sofue \etal 1987, Yusef-Zadeh \& Morris 1987a).
Adopting the middle of this range, $n_e B_\parallel L \approx 3000 \ut cm -3
\mu$G pc, we find $B_\parallel$ of 2 or 18 $\mu$G for L=100 or 1 pc
respectively, significantly less than the limit on the total field
strength.  This is consistent with the idea that the large-scale field
runs vertically through the plane near the Galactic center.
The projection of the field on the plane of the sky must be fairly uniform
so that the image of Sgr A$^*$ is not circularized by the cancellation of
different contributions along the line of sight.
Future observations of the scatter-broadened shape of
of  OH/IR stars lying in the inner 15$'$ of the
Galactic center are crucial in testing this picture.

It has been suggested that the wind from the IRS 16  cluster which lies
within the inner pc of the Galaxy may in fact be responsible for the
scattering medium. However the required size of the density
fluctuations is larger for screens closer to Sgr A$^*$, both because
they are thinner and because the scattering angles become larger.  At
one pc from Sgr A$^*$, electron density fluctuations of 2000 cm$^{-3}$ are
needed, whereas the mean density of the IRS 16 wind at 1 pc is only 10
cm$^{-3}$. Further, an entrained magnetic field will be stretched
parallel to the East-West line joining IRS 16 and Sgr A$^*$,
scatter-broadening Sgr A$^*$ orthogonal to the observed direction.

Although the deviation in the scaling of the size
of the minor axis from a $\lambda^2$ law may not be significant
it is worth noting that
a deviation would raise an interesting problem
because anisotropic turbulence should produce a shape that is
frequency-independent. That is, both major and minor axes should depend
quadratically on wavelength.
One possibility is that the intrinsic emission from Sgr A$^*$ is
contributing to the change in its apparent shape. This is
intriguing given the recent prediction that a one to two
million solar mass black hole
accreting at the  location of Sgr A$^*$
 should have an {\it intrinsic} size of several mas at cm
wavelengths (Melia, Jokipii and Narayanan 1992).
The apparent flux distribution is a convolution of the emission intrinsic
to Sgr A* and the angular broadening associated with scintillation.
Recent 3D hydrodynamical  simulations (Ruffert and Melia 1993),
suggest that
at longer wavelengths, we are sampling
the larger sized structure which follows the general bow-shock pattern, whereas
as the frequency increases, smaller scale
lengths corresponding to the more compact region of the infalling
plasma are sampled.  Since the flow becomes more radial as it falls in toward
the blackhole, then one should expect to see a decrease in the axial
ratio as a function of increasing frequency. The slope of the
minor axis in Figure 1 may give a hint in support of the above suggestion.
However, it  is
 clear that an improved measurement of the minor axis at $\lambda$0.7,
1.35,  20 and 90cm are crucial to confirm the change in
 shape with wavelength as well as to
give better estimates of the intrinsic source size and the shape of the
anisotropy.

{\bf {Acknowledgements}}:
This work was supported  by NASA grant NAGW-2518.

\medskip
\centerline{\bf References}
\ref
Alberdi, A., Lara, L., Marcaide, J.M., Elosegui, P., Shapiro, I.I. \etal
 1993, {\it A.A.}, in press.
\ref
Backer, D.C. 1988, in {\it AIP Conf. Proc. 174: Radio Wave Scattering in
the Interstellar Medium}, eds. J. Cordes, B. Rickett, D. Backer,
[AIP : New York], p111.
\ref
Backer, D.C. 1993, in {\it Nuclei of Normal Galaxies: Lesson from the
Galactic Center}, ed. R. Genzel and A. Harris, in press.
\ref
Backer, D. C., Zensus, J.A., Kellermann, K.I., Reid, M., Moran, J.M. and
Lo, K.Y. 1993, {\it Science}, in press.
\ref
Balick, B. and Brown, R.L. 1974, {\it ApJ, \bf 194}, 265.
\ref
Brown, R.L. and Lo, K.Y. 1982 {\it ApJ, \bf 253}, 108.
\ref
Cordes, J.M., Anathakrishnan, S., and Dennison, B. 1984, {\it Nature,
309}, 689.
\ref
Davies. R.D., Walsh, D. and Booth, R.S. 1976, {\it MNRAS, \bf 177}, 319.
\ref
Eckart, A.,  Genzel, R., Hoffmann, R., Sams, B.J., and Tacconi-Garman,
L.E. 1993, {\it ApJ, \bf 407}, L77.
\ref
Handa, T., Sofue, Y., Nakai, N., Hirabayashi, H., Inoue, M. 1987, {\it
Publ.Astr.Soc.Japan, \bf 39}, 709.
\ref
Higdon, J.C. 1984, {\it ApJ, \bf 285}, 109.
\ref
Higdon, J.C. 1986, {\it ApJ, \bf 309}, 342.
\ref
Inoue, M., Takahashi, T., Tabara, H., Kato, T., Tsboi, M. 1984, {\it
Publ. Astr.Soc.Japan, \bf 36}, 633.
\ref
Jauncy, D.L., Tziousmis, A.K., Preston, R.A., Meier, D.L., Batchelor, R.
 \etal 1989, {\it A.J., \bf 98}, 44.
\ref
Krichbaum, T.P., Zensus, J.A., Witzel, A., Mezger, P.G., Standke, K.J.
 \etal 1993, {\it A.A. \bf 274}, L37.
\ref
Lo, K.Y., Backer, D.C., Ekers, R.D., Kellermann, K.I., Reid, M. and
Moran, J.M. 1985, {\it Nature, \bf 249}, 504.
\ref
Lo, K.Y., Backer, D.C., Kellermann, K.I., Reid, M., Zhao, J.H.
\etal, 1993, {\it Nature, \bf 362}, 38.
\ref
Marcaide, J.M., Alberti, A., Bartel, N., Clark, T.A., Corey, B.E.
 \etal 1992, {\it A.A., \bf 258}, 295.
\ref
Melia, F. 1992, {\it ApJ., \bf 387}, L25.
\ref
Melia, F., Jokippi, R. and Narayanan, A. 1992, {\it ApJ, \bf 395}, L87.
\ref
Mezger, P.G., and Pauls, T. 1985, in {\it IAU  Symposium No. 84}, ed.
W.B. Burton, [Reidel : Dordrecht], p. 357.
\ref
Ruffert, M. and Melia, F. 1993, {\it ApJ.Letters}, submitted.
\ref
Schmidt, J. 1978, dissertation, Univ. Bonn.
\ref
Serabyn, E. and G\"usten, R. 1991, {\it A.A. \bf 242}, 736.
\ref
Sofue, Y., Reich, W., Inoue, M. and Seiradakis, J.H. 1987, {\it
Publ.Astron.Soc.Japan, \bf 39}, 95.
\ref
Tsuboi, M., Inoue, M., Handa, T., Tabara, H., Kato \etal 1986, {\it
A.J., \bf 92}, 818.
\ref
van Langevelde, H.J., Frail, D.A., Cordes, J.M. and Diamond, P.J. 1992,
{\it ApJ, \bf 396}, 686.
\ref
Yusef-Zadeh, F.  1993, in {\it Nuclei of Normal Galaxies: Lesson from the
Galactic Center}, ed. R. Genzel and A. Harris, in press.
\ref
Yusef-Zadeh, F. and Morris, M. 1987a, {\it  ApJ, \bf 322}, 721.
\ref
------------------------------ 1987b, {\it  A.J., \bf 94}, 1178.
\ref
Yusef-Zadeh, F., Morris, M. and Chance D. 1984, {\it Nature, \bf 310}, 557.
\ref
Zhao, J.H., Goss, W.M., Lo, K.Y. and Ekers, R.D. 1991, in {\it
Relationship between Active Galactic Nuclei and Starburst Galaxies}, ed.
A. Fillipinko, [ASP: San Fransisco], p295.

\medskip
\centerline{\bf Figure Caption}

{\bf Figure 1.} A logarithmic plot showing  the size of the semi-major
(top) and semi-minor (bottom) axes of Sgr A$^*$ at a number of frequencies
(Backer \etal 1993; Alberdi \etal 1993; Lo \etal 1993; this paper). The
fit to each axis is also shown.

\vfill\eject\null
\pageno 12\hbox{\bf AUTHORS' ADDRESS:}
\vskip 0.3 in
\settabs\+\noindent&{\bf Farhad Yusef-Zadeh:}\quad&\cr
\+&{\bf William Cotton:}&National Radio Astronomy Observatory,\cr
\+&\null&Edgemont Road, Charlottesville, VA 22903\cr
\+&\null&[cotton@nrao.edu]\cr
\+&{\bf Fulvio Melia:}&Department of Physics and Steward Observatory,\cr
\+&\null&University of Arizona, Tucson, AZ 85721\cr
\+&\null&[melia@nucleus.physics.arizona.edu]\cr
\+&{\bf Douglas A. Roberts:}&Department of Astronomy,\cr
\+&\null&University of Illinois, Urbana, IL 61801\cr
\+&\null&[droberts@sirius.astro.uiuc.edu]\cr
\+&{\bf Mark Wardle:}&Department of Physics and Astronomy,\cr
\+&\null&University of Rochester, Rochester, NY 14627-0011\cr
\+&\null&[mark@holmes.astro.nwu.edu]\cr
\+&{\bf Farhad Yusef-Zadeh:}&Department of Physics and Astronomy,\cr
\+&\null&Northwestern University, Evanston, IL 60208\cr
\+&\null&[zadeh@ossenu.astro.nwu.edu]\cr
\end